\documentclass[apjl]{emulateapj} 
\usepackage{amsmath}
\usepackage{natbib}
\usepackage{graphicx}
\usepackage{epsf}
\usepackage{color}
\DeclareGraphicsExtensions{.jpg,.pdf,.png,.eps,.ps}
\graphicspath{{FIGURES/}}

\newcommand{\planck}{{\it Planck}}

\newcommand{\ltsima}{$\; \buildrel < \over \sim \;$}
\newcommand{\ltsim}{\lower.5ex\hbox{\ltsima}}

\newcommand{\omegae}{\Omega_{\rm e}}
\newcommand{\od}{\Omega_{\rm de}}
\newcommand{\beq}{\begin{equation}}
\newcommand{\eeq}{\end{equation}}
\newcommand{\beqa}{\begin{eqnarray}}
\newcommand{\eeqa}{\end{eqnarray}}

\def\ha{\frac{1}{2}}

\begin{document}

\title{New limits on Early Dark Energy from the South Pole Telescope}

\author{
 C.~L.~Reichardt,\altaffilmark{1}
 R.~de~Putter,\altaffilmark{2,3}
  O.~Zahn,\altaffilmark{4} and 
  Z.~Hou\altaffilmark{5}
}

\altaffiltext{1}{Department of Physics,
University of California, Berkeley, CA USA 94720}
\altaffiltext{2}{IFIC, Universidad de Valencia-CSIC, 46071, Valencia, Spain}
\altaffiltext{3}{ICC, University of Barcelona (IEEC-UB), Marti i Franques 1, Barcelona 08028, Spain}
\altaffiltext{4}{Berkeley Center for Cosmological Physics,
Department of Physics, University of California, and Lawrence Berkeley
National Labs, Berkeley, CA, USA 94720}
\altaffiltext{5}{Department of Physics, 
University of California, Davis, CA, USA 95616}

\email{cr@bolo.berkeley.edu}
 
\begin{abstract}

We present new limits on early dark energy (EDE) from the cosmic microwave background (CMB) using data from the WMAP
satellite on large angular scales and South Pole Telescope (SPT) on small angular scales. 
We find a strong upper limit on the EDE density of $\Omega_e < 0.018$ at 95\% confidence, a factor of three improvement over WMAP data alone.  
We show that adding lower-redshift probes of the expansion rate to the CMB data improves constraints on the
dark energy equation of state, but not the EDE density.
We also explain how  small-scale CMB temperature anisotropy constrains EDE. 

\end{abstract}


\bigskip\bigskip

\section{Introduction}

The mystery of cosmic acceleration is one of the most pressing questions in cosmology. 
We have strong evidence that the expansion of the Universe is accelerating from observations of supernovae (SNe --
\cite{perlmutter99a,riess98,amanullah10}),
the cosmic microwave background (CMB -- \cite{komatsu11,das11b,keisler11}) and galaxy surveys (\cite{reid10b,blakeetal11}). 
The simplest model for this acceleration is a cosmological constant. 
However, a wide range of alternative explanations exist, from modifications to general relativity to scalar fields. 

To explain the observed cosmic acceleration, the dark energy (DE) equation of state (e.o.s.) must be close
to $w = -1$ at late times. If $w$ is constant, as for a cosmological
constant and certain quintessence models, DE quickly becomes irrelevant as one looks back in cosmic history  (the cosmological constant would be  $\sim\,10^{-9}$ of the total density at the time of CMB last scattering).
In early dark energy (EDE) models, the e.o.s.~was instead much larger in the past
so that the DE density was considerable even in the early universe.

The best motivated EDE theories are so called ``tracker'' models
(e.g.~\cite{ratrapeebles,PeebRa88, wett88,hebwett01}), 
which help to ameliorate the coincidence problem.
While in non-early dynamical DE models
the initial conditions must be very finely tuned in order for the DE density to be of the same order of magnitude
as the matter density today, tracker models contain attractor solutions on which the DE evolution is
determined by the e.o.s.~of the dominant component of the universe, i.e.~matter or radiation.
The advantages are that the DE density can stay much closer to the energy density of the dominant component
and that the DE evolution is independent of initial conditions. 
Tracer models, a subclass of tracker models, further limit  the DE e.o.s.~to be equal to the background e.o.s.~so that $w=1/3$
during radiation domination and $w=0$ during matter domination. While the EDE fraction in generic tracker models
is typically still quite small (although larger than in the constant $w$ case), in tracer models
it can easily be at the several percent level.

Several groups have used observations of the CMB, baryon acoustic oscillations (BAO), SNe, Ly-$\alpha$ forest and Hubble constant
to constrain EDE models \citep{DorRob06,xia09, alam11, calabrese11}. 
Current observations show no preference for a non-zero EDE density. 
The  best upper limit on the EDE density from the CMB today is by \citet{calabrese11}. 
Using large and small angular scale CMB observations from the WMAP7, ACBAR, QUaD and ACT experiments \citep{komatsu11, reichardt09a, gupta10, das11b}, they find a 95\% confidence upper limit on the EDE density of $\Omega_e < 0.043$ for a ``relativistic" EDE model and $\Omega_e < 0.024$ for a ``quintessence" model. 
As demonstrated by \citet{xia09}, adding high redshift measures of the growth factors and matter power spectrum can lead to much tighter (if potentially modeling dependent) constraints on EDE. 

In this work, we present EDE constraints derived from the latest measurements of small-scale CMB anisotropy. 
We use bandpowers from the South Pole Telescope (SPT) \citep[hereafter K11]{keisler11}, which are the best measurement of the CMB across the third to ninth acoustic peaks. 
We combine the SPT bandpowers with the WMAP7 bandpowers on larger angular scales in order to significantly tighten upper limits on the allowed EDE density. 
We also consider the effect of adding BAO and SNe data. 

We present the EDE model  in \S\ref{sec:edemodel}, and detail how EDE affects small-scale CMB anisotropy  in \S\ref{sec:intuit}. 
We describe the data we use in \S\ref{sec:data}. Results are presented in \S\ref{sec:results}, and we conclude in \S\ref{sec:conclusion}. 

\section{EDE Model}
\label{sec:edemodel}

Instead of considering a specific model, we choose to constrain EDE  in the
more general  tracer model parametrization by \cite{DorRob06}. 
This parametrization introduces two new parameters on top of the $\Lambda$CDM set,
the DE e.o.s.~at $z=0$, $w_0$, and the DE density
relative to the critical density at early times, $\Omega_e$, which is taken to be constant at sufficiently high redshift ($z \gtrsim 10$). 
The DE density and e.o.s.~are then given by
\beqa
\od(a) &=&  \frac{\od^0 - \omegae \left(1- a^{-3 w_0}\right) }{\od^0 + \Omega_{m}^{0} a^{3w_0}} + \omegae \left (1- a^{-3 w_0}\right)\\
w(a) &=& -\frac{1}{3[1-\od(a)]} \frac{d\ln\od(a)}{d\ln a} + \frac{a_{eq}}{3(a + a_{eq})}. 
\eeqa
Here $a_{eq}$ is the scale factor at matter-radiation equality, 
and $\od^0$ ($\Omega_{m}^{0}$) is the DE (matter) density relative to
critical density at $z=0$. 
This parametrization assumes spatial flatness so that $\Omega_{m}^{0} + \od^0 = 1$.
Since we force $\Omega_{\rm de}(a)$ to be constant at high redshift, the DE e.o.s.~mimics that of the dominant component at early times, thus behaving like a tracer model.
Later during matter domination, at $z < 10$, the e.o.s.~transitions towards
its current value, $w_0$, so it can account for cosmic acceleration.

To consistently describe the perturbations, we are motivated by quintessence models
to treat DE as a perfect fluid with a sound
speed, $c_s$, equal to the speed of light (see \cite{Hu98}). 
This choice together with the parametrization for the background evolution completely
specifies the behavior of DE. 
We only consider models
with $w_0 \geq -1$ and thus do not entertain the possibility of
``phantom crossing'' (see e.g.~\cite{fanghulew08}). 
This restriction allows us to avoid pathologies
in perturbation evolution and to stay in the quintessence regime.

\section{How the CMB constrains EDE}
\label{sec:intuit}

EDE directly  influences the evolution of acoustic oscillations, 
 and thus imprints a much stronger signature on the CMB temperature anisotropy than late-time DE. 
In contrast, late-time DE only affects the CMB through its effect on the distance to
last scattering and the integrated Sachs Wolfe (ISW) effect, leading to a strong degeneracy  between DE parameters with CMB data alone.

First of all, EDE increases the expansion rate, rescaling the
Hubble parameter at high redshift by a constant factor $H \to \left(1 - \Omega_e\right)^{-\ha} \, H$.
Since quintessence-like (with $c_s = 1$, the speed of light) DE does not cluster on scales smaller than the horizon,
the result is a suppression of the growth of cold dark matter and metric perturbations after matter radiation equality.
This suppression in turn drives a boost in the amplitude of oscillations in the baryon-photon plasma. 

The effect of adding EDE is similar to the better known effect of increasing the radiation to matter ratio (see e.g.~\cite{hu97b}). 
Indeed, \citet{calabrese11} show that EDE can be related to a time-dependent effective number of relativistic species, $\Delta N_{\rm eff}$, which is roughly four times higher at recombination than during big-bang nucleosynthesis (BBN). 
Temperature modes entering during radiation domination ($\ell > \ell_{\rm eq}$) when the growth of metric perturbations
is suppressed, undergo a boost relative to those entering during matter domination ($\ell < \ell_{\rm eq}$). 
Increasing the radiation to matter ratio shifts $\ell_{\rm eq}$ to lower values, thereby
enhancing the amplitude at $\ell$ near $\ell_{\rm eq}$. 

Analogously, $\Omega_e$ shifts
the scale corresponding to the time of equality between clustering matter and non-clustering radiation \&
DE according to $\ell_{\rm eq} \to \sqrt{1 - 2 \Omega_e} \, \ell_{\rm eq}$ (since the distance
to the CMB last scattering surface also decreases with increasing $\Omega_e$, EDE causes the angular
mode $\ell_{\rm eq}$ to shift to a smaller value).
However, the exact signature of the boost has a different scale dependence than the effect due to
simply changing the relative amounts of matter and radiation (for example by varying the effective number of
relativistic species, $N_{\rm eff}$) because, after radiation domination, the DE density does not decay as fast as that
of radiation.

\begin{figure*}[htb!]
\centering
\includegraphics[width=0.9\textwidth]{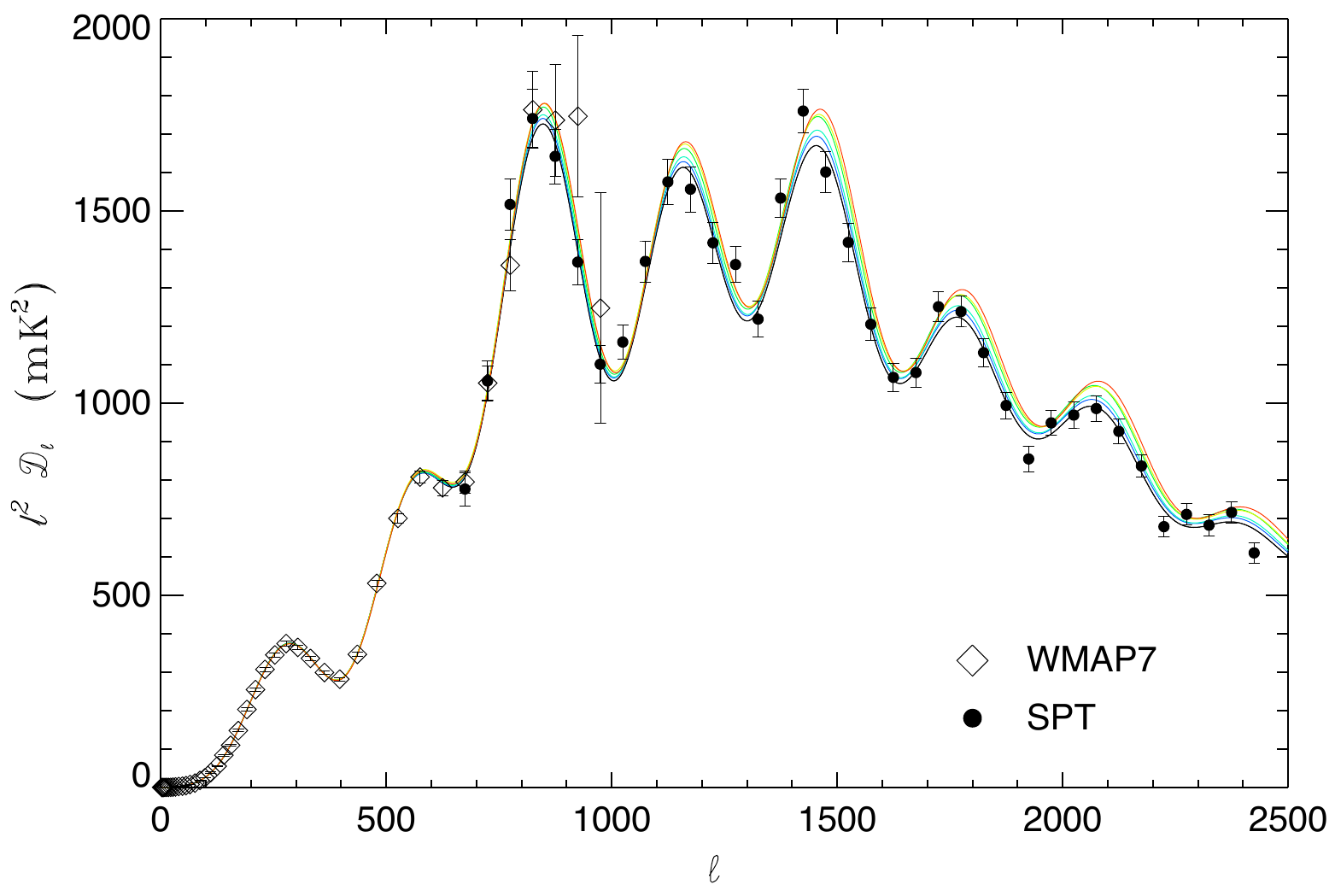}
 \caption{ 
The WMAP7 bandpowers over-plotted on the best-fit WMAP7 model as $\Omega_e$ varies from 0 (black) to 0.05 (red). 
  All the models can accommodate the WMAP7 bandpowers at low $\ell$. 
  However, models with more EDE predict more power at higher $\ell$ (smaller angular scales). 
  Measuring the CMB power spectrum on small angular scales with experiments such as SPT (and \planck\ in the future) can thereby strengthen the limits on EDE. 
We include the SPT bandpowers in the plot to illustrate their constraining power. 
  }
  \label{fig:highl-intuit}
\end{figure*}

In addition to boosting the acoustic oscillations, the increase in the Hubble scale due to EDE changes the sound horizon scale at decoupling, the Silk damping scale
and the distance to last scattering. Finally, the lack of clustering in the DE contributes to
the early ISW effect, and hence increases power for modes larger than the first acoustic peak.

While the signatures discussed above are to a degree degenerate with variations in other cosmological
parameters, there remains a clear EDE signature even after taking this into account,
as shown in Fig.~\ref{fig:highl-intuit}. 
Here, for values of $\Omega_e$ spaced in 0.01 increments from $\Omega_e = 0$ to 0.05, we show the best-fit WMAP7 \citep{larson10} spectrum.
Clearly, after accounting for
parameter degeneracies with the large-scale CMB data, there is a strong signature of EDE on small scales ($\ell > 1000$). 
Increasing the EDE density increases the small scale power  because it shortens the Silk damping length at the last scattering surface. 
The angular diameter distance decreases to a lesser degree; somewhat reducing the net change to the angular size of Silk damping. 
The physical sound horizon and angular diameter distance are reduced by similar amounts so the positions of the acoustic peaks move only slightly to larger scales. 
The reduced damping further reduces the apparent shift in position of the first peaks. 
The clear signal at $\ell > 1000$ strongly motivates using the SPT power spectrum measurement by K11, as it provides
the best measurement  to date of the CMB temperature anisotropy over the Silk damping tail. 

EDE also affects the BBN helium prediction; however this effect does not contribute significantly to the current CMB constraints. 
With respect to BBN, EDE is equivalent to
adding (non-interacting) relativistic species, $\Delta N_{\rm eff} \sim 7.44 \Omega_e$, which
affect the helium fraction through $\Delta Y_{\rm He} \sim 0.013 \Delta N_{\rm eff}$ \citep{calabrese11}.
Even $\Omega_e = 0.05$ (the $2\,\sigma$ limit from WMAP7 only) would change the helium fraction by only $\Delta Y_{\rm He} \sim 0.005$, which is small compared to the WMAP7+SPT $Y_{\rm He}$ uncertainty of $0.03$ (K11). 
The impact of EDE on the helium abundance is therefore negligible for the CMB constraints.

There are clear reasons to concentrate on the CMB anisotropy when considering EDE theories. 
Given that we have implemented a generalized tracer model for the EDE, the model mimics normal DE in the current, DE-dominated epoch. 
The effect of $\Omega_e$ at late times is similar to having a varying e.o.s.~$w_a \approx 5  \Omega_e$ (\cite{linrob08}) and is thus very weak.
Therefore, observations of the low-redshift universe, such as SNe, BAO, and CMB lensing,
primarily constrain $\od^0$ and $w_0$.

\section{Data}
\label{sec:data}

This work focuses on how small-scale CMB data complement large-scale CMB data to constrain EDE. 
Therefore we include large-scale CMB bandpowers from the seven-year WMAP data release (WMAP7, \citealt{larson10}) in all parameter chains. 
The best current constraints on small-scale CMB power spectrum come from an analysis of $800\,{\rm deg}^2$  of sky observed with the SPT (K11). 
The K11 bandpowers cover angular multipoles from $500 < \ell < 3000$.
The covariance between the WMAP and K11 bandpowers is negligible (K11). 
However, it is non-trivial to combine the K11 bandpowers with those from other experiments focused on the CMB damping tail (e.g.~\citet{reichardt09a,das11}) since the experiments are cosmic variance limited and have overlapping sky coverage. 
Hence we restrict ourselves to the K11 bandpowers to constrain the small-scale CMB temperature anisotropy.

We also consider EDE constraints from geometrical observations at lower redshift. 
Specifically, we look at three additional datasets: BAO observations, Hubble constant data (referred to as `H$_0$'), and  SNe data.
We use observations from SDSS and 2dFGRS of the BAO feature \citep{percival10}. 
\citet{percival10} use the BAO feature to measure the angular diameter distance relation. 
We also use direct, low-redshift measurements of the Hubble constant with the Hubble Space Telescope that found $H_0 = 73.8 \pm 2.4\,$km\,s$^{-1}$\,Mpc$^{-1}$ \citep{riess11}.
For SNe, we add the Supernovae Cosmology Project's Union2 dataset \citep{amanullah10} which constrains the luminosity distance-redshift relation.  

\section{Results}
\label{sec:results}

We fit the data to a model for the lensed primary CMB anisotropy. 
Parameter constraints are calculated using the publicly available {\textsc CosmoMC}\footnote{http://cosmologist.info/cosmomc (Aug 2011)} package \citep{lewis02b}. 
The CMB power spectrum for a given cosmology is calculated using the CAMB\footnote{http://camb.info (July 2011)} package \citep{lewis00}. 
We have modified both packages to include the EDE prescription outlined in \S\ref{sec:edemodel}.  
This model has eight free parameters: $\Omega_b h^2$, $\Omega_c h^2$, $A_s$, $n_s$, $\tau$, $\Omega_{\rm de}^0$, $w_0$ and $\Omega_e$. 
We do not consider additional cosmological parameters such as the neutrino energy density fraction or $N_{\rm eff}$,
but note that \cite{joudakikap11} and \cite{calabrese11} suggest that $\Omega_e$ is largely independent of these parameters. 
We also include in all parameter chains the three nuisance foreground terms described by K11; these nuisance terms negate the information from the amplitude (but not peak structure) of the power spectrum at $\ell \gtrsim 2500$. 
We have made one change to the foreground terms; we have changed the template shape of the clustered component of the cosmic infrared background to a pure power law ($D_\ell \propto \ell^{0.8}$) to reflect recent constraints from SPT, ACT, Planck, and BLAST \citep{reichardt11a, addison11}. 
The Monte Carlo Markov chains are available for download at the SPT website.\footnote{http://pole.uchicago.edu/public/data/ede11/}

\begin{figure}[t]\centering
\includegraphics[width=\columnwidth]{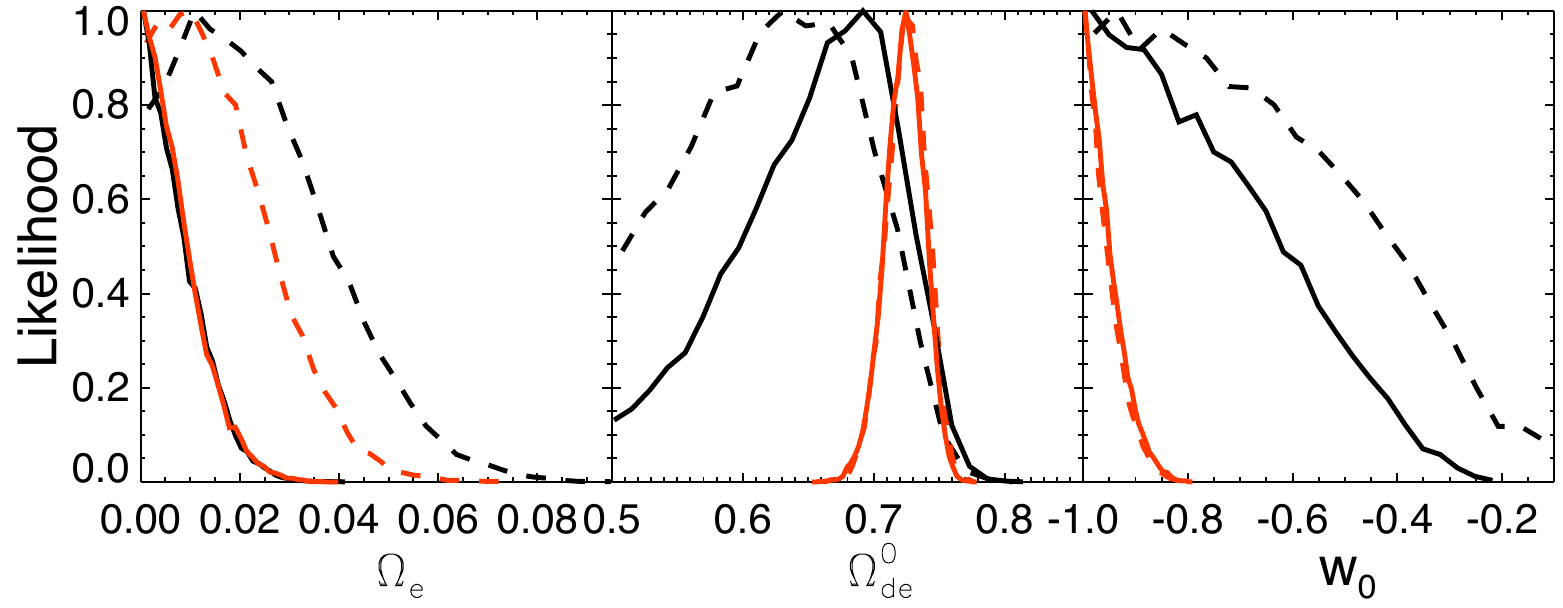}
  \caption[]{ 
  Posterior probabilities for $\Omega_e$ (\textbf{\textit{Left panel}}), $\Omega_{\rm de}^0$ (\textbf{\textit{Middle panel}}), and $w_0$ (\textbf{\textit{Right panel}}).
  {\bf Black lines} show constraints from the CMB alone. 
   {\bf Red lines} show constraints from the CMB + BAO + H$_0$ + SNe.
   {\bf Solid lines} include the SPT bandpowers from K11 in addition to WMAP7; {\bf dashed lines} include only WMAP7 data. 
    }
  \label{fig:likeede}
\end{figure}

We consider the eight data combinations described in \S\ref{sec:data}. 
The resulting constraints on DE are summarized in Table \ref{tab:ede} and plotted in Figure \ref{fig:likeede}. 
With only the large-scale CMB observations from WMAP, the geometrical measures tighten the upper limit on EDE by 30\% from $\Omega_e < 0.052$ to $\Omega_e < 0.037$ at 95\% CL. 
This improvement results from better measurements of $\Omega_mh^2$, $n_s$ and $\Theta_s$, which are somewhat degenerate with  $\Omega_e$. 
Notably however, as expected from \S\ref{sec:intuit}, the EDE constraints are completely determined by the CMB power spectrum once the spectrum is measured over the Silk damping tail. 
The small-scale CMB bandpowers tighten the upper limit by a factor of three over the large-scale CMB data alone to $\Omega_e < 0.018$. 
The 95\% confidence upper limit on $\Omega_e$ remains unchanged at $\sim$0.018 when  BAO,  H$_0$  and SNe observations are added to the CMB power spectrum.
 
Conversely, the geometrical constraints are essential for the local (z=0) DE parameters. 
Both the current DE density and its e.o.s.~are poorly constrained by the CMB alone, as shown in Table \ref{tab:ede} and Figure \ref{fig:likeede}.  

\begin{figure}[thbp]\centering
\includegraphics[width=\columnwidth]{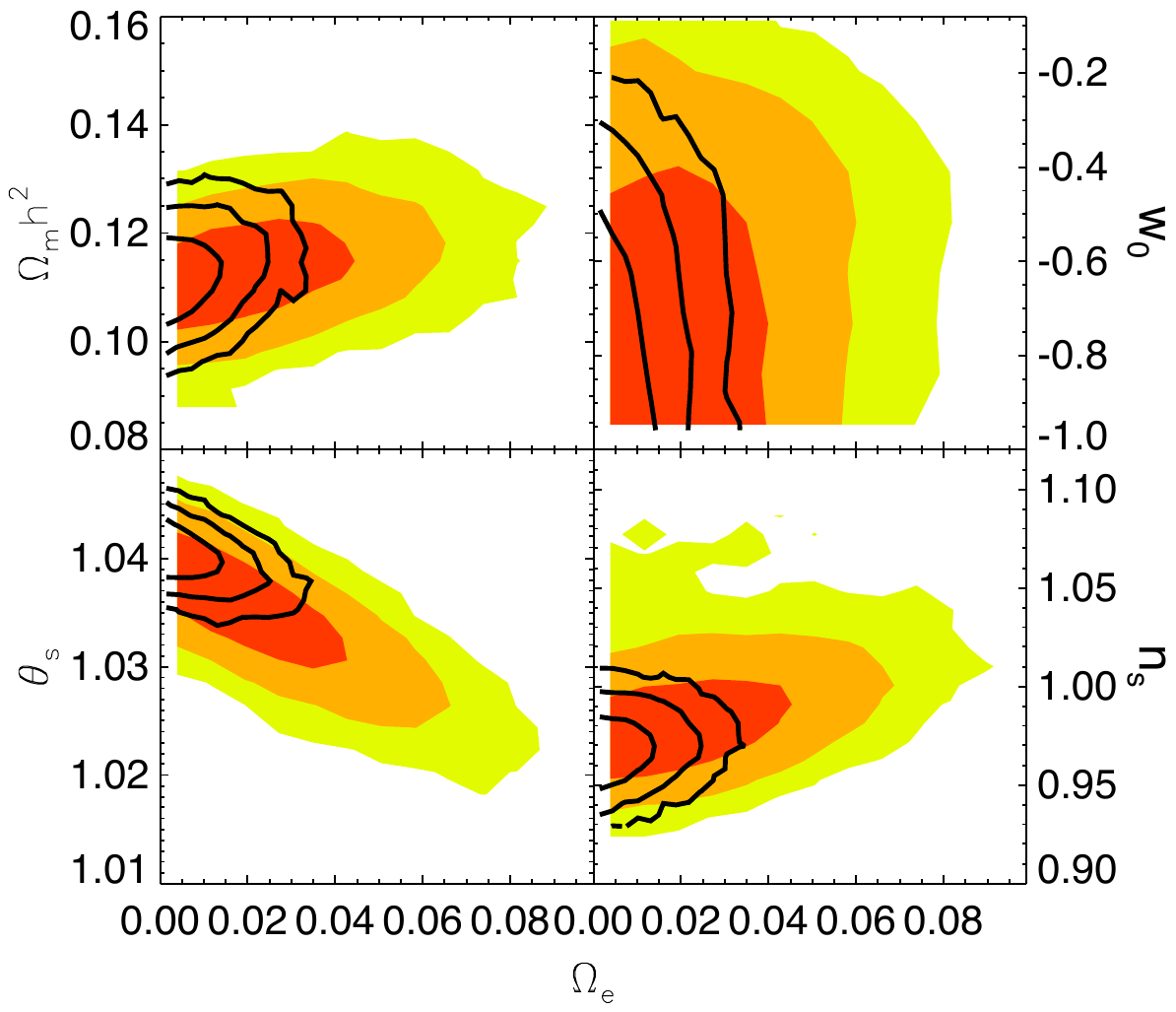}
  \caption[]{ 
  2D likelihood surfaces between $\Omega_e$ and the three most correlated parameters: $\Omega_m h^2$ ({\it top left panel}), $\Theta_s$ ({\it bottom left panel}), and $n_s$ ({\it bottom right panel}). 
  We also show the likelihood surface for $\Omega_e$ and the DE e.o.s.~at $z=0$, $w_0$ ({\it top right panel}). 
  $\Omega_e$ is largely independent of the remaining parameters ($\Omega_{\rm de}^0$, $A_s$, $\tau$ and $\Omega_b h^2$). 
  Contours mark the $68.3, 95.4$ and $99.7 \%$ confidence regions. 
  The {\bf filled, colored contours} show the constraints from WMAP7 data only. 
  The {\bf black line contours} show the constraints when the SPT small-scale CMB bandpowers are added.
  }
    \label{fig:like2d}
\end{figure}

We show the main degeneracies in the CMB data between the EDE density and other parameters in Figure \ref{fig:like2d}. 
The red, orange and yellow filled contours denote the $68.3, 95.4$ and $99.7 \%$ CL regions with WMAP7 data. 
The solid black lines mark the same likelihood regions for WMAP7 + SPT data. 
We do not see a strong degeneracy between the EDE density and either late time DE parameter ($w_0$ or $\Omega_{\rm de}^0$). 
The $w_0$-$\Omega_e$ plane is shown in the top right panel. 
We do see degeneracies in the WMAP7 constraints between $\Omega_e$ and $\Omega_m h^2$, $\Theta_s$, and $n_s$, shown in the top left, bottom left and bottom right panels respectively. 
Perhaps unsurprisingly given the importance of the high-$\ell$ bandpowers to measuring EDE, these are also the three parameters for which adding small-scale CMB data has the largest impact (K11). 
The strongest degeneracy is between $\Omega_e$ and the angular scale of the acoustic peaks, $\Theta_s$.  
This is  the $\Lambda$CDM  parameter  best constrained by the SPT data since the SPT bandpowers improve measurements of the third to ninth acoustic peaks. 
The SPT bandpowers also improve the WMAP7 constraints on $n_s$ by 30\% and on $\Omega_m h^2$ by 15\%. 
The $\Omega_m h^2$ is coming primarily from the third acoustic peak, while the $n_s$ constraint comes from extending the range of angular scales probed. 
For all three parameters, the SPT bandpowers reduce the degeneracies with $\Omega_e$. 
As shown in Figure  \ref{fig:like2d}, the allowed regions of the parameter planes are reduced by much more than the 15-30\% numbers mentioned above.

\begin{table*}[hbt!]
\begin{center}
\caption{\label{tab:ede} Dark energy constraints}
\small
\begin{tabular}{c|cccc}
\hline\hline
\rule[-2mm]{0mm}{6mm}
$\Omega_e$  & CMB & CMB + BAO + H$_0$ & CMB + SNe & CMB + BAO + H$_0$ +SNe \\
\hline
WMAP7 only& $<0.052$ & $<0.038$ & $<0.039$& $<0.037$ \\
\smallskip
WMAP7+SPT& $<0.018$ & $<0.018$& $<0.018$& $<0.019$ \\
$\Omega_{\rm de}^0$  & &&& \\
\hline
WMAP7 only&$0.608 \pm 0.093$ & $0.720\pm 0.018$ &$0.716 \pm 0.025$ &$0.726\pm 0.016$ \\
\smallskip
WMAP7+SPT& $0.660\pm0.067$& $0.720\pm 0.017$&$0.722\pm 0.023$ & $0.724\pm 0.015$ \\

$w_0$  &  &&& \\

\hline
WMAP7 only&$<-0.29$ &$<-0.84$  &$<-0.87$ & $<-0.89$\\

WMAP7+SPT&$<-0.45$& $<-0.84$& $<-0.86$& $<-0.90$ \\

\hline
\end{tabular}
\tablecomments{Constraints on the early and late-time ($z=0$) DE densities, $\Omega_e$ and $\Omega_{\rm de}^0$, and e.o.s.,
$w_0$, for different datasets. 
We report 95\% confidence upper limits on the EDE density and DE e.o.s.. 
The priors on these parameters require that they be greater than 0 and -1 respectively. 
We report $1\,\sigma$ uncertainties on the $z = 0$ DE density, $\Omega_{\rm de}^0$. 
Geometrical constraints from BAO, H$_0$ and SNe are very important for the late-time DE parameters ($w_0$ and $\Omega_{\rm de}^0$). 
However, the constraint on the EDE density is primarily from the CMB temperature anisotropy. 
} \normalsize
\end{center}
\end{table*}

The effect of EDE on observables depends on its clustering properties.
As discussed in section \ref{sec:edemodel}, inspired by quintessence models, the best motivated approach is to treat
the DE as a perfect fluid with rest frame sound speed $c_s = 1$.
However, more exotic models such as k-essence (\cite{armmukstein00,chiokyam00}) can have sound speeds below the speed of light
and thus allow significant DE perturbations on sub-horizon scales. 
To illustrate how the EDE bound depends on assumptions about DE perturbations,
we briefly consider two alternative models with $c_s = 0$ or $1/\sqrt3$. 
In the $c_s = 0$ case, DE is free to cluster
on all scales, and in this particular sense it behaves more like cold dark matter,
although the DE perturbations receive the usual suppression when the e.o.s.~$w \to -1$ at late times
(e.g.,~\cite{rdphutlin10}).
Our default $c_s=1$ case is more similar to radiation where perturbation growth is suppressed by pressure perturbations. 
We also look at $c_s = 1/\sqrt3$ since it leads to weakest constraints on $\Omega_e$. 

In the $c_s = 0$ case, the WMAP7 upper limit on the EDE density  is $\Omega_e < 0.049$, nearly unchanged from the baseline $c_s=1$ case. 
However, the limits weaken for the other data sets. 
The 95\% upper limit strengthens to $\Omega_e < 0.035$ with WMAP7 + K11, and remains nearly unchanged  for the maximal data set WMAP7 + K11+ BAO + H$_0$ +SNe. 
These limits are a factor of two weaker than in the baseline case.

Looking at the $c_s$-$\Omega_e$ plane, we find the weakest constraints on $\Omega_e$ at $c_s^2 = 1/3$. 
For WMAP7 only, the 95\% upper limit is $\Omega_e < 0.119$, a factor of two higher than for $c_s^2 = 1$. 
Adding BAO + H$_0$ +SNe or K11 data improves the limit by a factor of 1.6. 
Adding all of these reduces the 95\% upper limit to $\Omega_e < 0.055$,  a factor of three higher than for $c_s^2 = 1$.

\section{Conclusion}
\label{sec:conclusion}
We find an upper limit on the EDE density of $\Omega_e < 0.018$ from CMB data alone. 
We show that the SPT data make observations of the low redshift expansion history redundant, i.e. the CMB constraint is not improved further using BAO or SNe data. 
The constraining power comes from EDE's impact on the Universe's expansion rate at $z \sim 1100$. 
This leads to relatively more power at small angular scales and to small shifts in the peak positions. 
Small angular scale CMB observations of the higher acoustic peaks break degeneracies in the WMAP data between the EDE density and the position of the acoustic peaks, the matter density and slope of the primordial power spectrum. 
In the next year, we expect order of magnitude improvements in measurements of the small-angular scale CMB power spectrum from the SPT, ACT and especially \planck\ surveys. 
These experiments promise to further elucidate the properties of DE in the early Universe. 

\acknowledgments

We thank Eric Linder, Lloyd Knox, and Adrian Lee for useful discussions.
This work was supported by the NSF through grants ANT-0638937 and ANT-0130612.  
RdP is supported by FP7-IDEAS-Phys.LSS 240117. 
This research used resources at NERSC, which is supported by the DOE. 
We acknowledge the use of LAMBDA which is supported  by NASA.



\end{document}